\documentclass[prb,twocolumn,showpacs]{revtex4}

\usepackage{amsmath}
\newcommand{\bs}{\boldsymbol}
\newcommand{\ra}{\rangle}
\newcommand{\la}{\langle}
\newcommand{\nn}{\nonumber}
\newcommand{\sn}{\section}
\newcommand{\h}{\hspace}
\newcommand{\lf}{\left}
\newcommand{\rt}{\right}
\newcommand{\be}{\begin{equation}}
\newcommand{\ee}{\end{equation}}
\newcommand{\bearr}{\begin{eqnarray}}
\newcommand{\eearr}{\end{eqnarray}}
\newcommand{\bpm}{\begin{pmatrix}}
\newcommand{\epm}{\end{pmatrix}}
\newcommand{\sig} {\sigma}
\newcommand{\veps}{\varepsilon}
\def \mrm {\mathrm}
\newcommand { \nvec } [1 ] { \boldsymbol {#1}}
\newcommand { \hvec } [1 ] { \boldsymbol{\hat #1}}
\newcommand{\btimes}{\boldsymbol \times}

\begin{document}

\title [Understanding SHE in 2DFS with generic SOI]
{Understanding spin Hall effect in 
two-dimensional fermionic systems with generic spin-orbit interaction in
III-V heterostructures}
%\shorttitle{Understanding SHE in 2DFS with generic SOI}
%Insert here a short version of the title if it exceeds 70 characters

\author{
Boudhayan Paul and Tarun Kanti Ghosh}

\affiliation{
Department of Physics, Indian Institute of Technology Kanpur
- Kanpur 208 016, India
}

\begin{abstract}
We delve into spin Hall effect in generic spin-orbit coupled
two-dimensional fermionic systems.
We derive analytically the spin-orbit force responsible for the 
spin Hall effect, and find that it has `Lorentz force'-like form.
We also derive the pseudo magnetic field responsible for this force.
We establish the relation between the
spin Hall conductivity, flux quanta and
this pseudo magnetic field, similar to
the one between charge Hall conductivity, flux quanta and the external field. 
We also present an exact closed-form expression
of the spin Hall conductivity in a generic spin-orbit coupled system.
%Depending on the dimensionality of the spin-orbit interactions, 
%the spin Hall conductivity depends on different combinations of
%Rashba parameter and fermion density.
\end{abstract}


\date{\today}

\maketitle

\sn{Introduction}

The (charge) Hall effect \cite {ashcroft}, which exists due essentially
to the Lorentz force, 
is a well known phenomenon of condensed matter physics. 
Similar to the charge Hall effect, there was a theoretical proposal that 
dc electric field can generate a spin Hall current in
electron/hole gases in III-V zinc blende semiconductors such as
AlGaAs-GaAs
\cite{zhang-03,zhang-04-1,niu-04-2,loss-04,sinova-04}.
This effect is also called intrinsic spin Hall effect (SHE)
where  spin accumulates on the edges parallel to the external 
electric field. The directions
of the accumulated spins are opposite on the opposite edges.
It is similar to the charge Hall effect, where charges of
opposite sign appear on the opposite edges.
The main difference is that contrary to the charge Hall effect, there is 
no need to apply any magnetic field.
A similar effect known as extrinsic spin Hall effect had been predicted by
Dyakonov-Perel \cite {perel-71-1,perel-71-2}
and Hirsch  \cite{hirsch-99} long ago.
It is termed {\it extrinsic} as it necessarily requires
spin dependent scattering from  magnetic impurities \cite{ahe,ahe1}.
In contrast, the {\it intrinsic} spin Hall effect is due entirely
to spin-orbit interaction (SOI) and occurs even in the absence 
of any scattering process.

Two independent groups have demonstrated experimental evidences of
the intrinsic spin Hall effect \cite{kato-04,sinova-05}.
It is believed that the intrinsic spin Hall effect has been realized in
2D spin-orbit coupled heavy hole systems by optical means. 
The observation of spin accumulation
established that there is a flow of pure spin current transverse to
an external electric field.
A pure spin current is thought of as a combination of
a current of spin-up electrons in one direction and current
of spin-down electrons in the opposite direction, resulting
in a flow of spin angular momentum with no net charge current.
Thus the intrinsic spin Hall conductance 
cannot be obtained by (charge) current measurements.

Although there have been numerous studies
\cite{murakami-04,berenvig,shen-04,culcer,loss-05,huang-06,ijmpb,eugen,
slf,njp}
on different aspects
of SHE in various condensed matter systems,
not many of them attempt to understand the origin of SHE.
Shen \cite{berry} has shown that the spin Hall conductivity is directly related to
the Berry phase.
%In Refs. \cite{sof1,sof2} the authors
Li \cite{sof1} and Nikolic \cite{sof2} \textit{et al} have tried to explain
SHE by deriving the pseudo magnetic field from
Lorentz-like spin-orbit force only in a Rashba 
coupled 2D electron gas.
But the study was incomplete as Lorentz force
involves velocity which, for spin-orbit coupled systems,
is not simply proportional to the (crystal) momentum
but involves additional SOI terms. In our investigation, we calculate
the spin-orbit force with the correct velocity expression,
derived in the context of two-dimensional fermionic systems
with generic spin-orbit interaction. We observe that this force
has a Lorentz-like form, which consistently explains the flow
of opposite spins in opposite directions.
Then we seek to extract the corresponding ``magnetic field''
for this Lorentz-like spin-orbit force.
We know that charge Hall conductivity is inversely
proportional to the external magnetic field.
We seek to find out the existence of a similar relation between spin
Hall conductivity (SHC) and this pseudo magnetic field.
In the later part of this work, we calculate the spin Hall
conductivity and establish an inverse square root relation with
the spin-orbit (magnetic) field. This relation also involves the unit
of (magnetic) flux quanta, which is significant from the fundamental
point of view. Our results also indicate
the dependence of SHC on spin-orbit coupling (SOC) constant
and fermion density through the higher order terms.

This report is organized as follows. In section II, 
we briefly mention the generic spin-orbit coupled Hamiltonian, 
its energy eigenvalues and the corresponding  eigenfunctions.
In section III, we derive Lorentz-like spin-orbit force and extract 
the spin-orbit interaction dependent pseudo magnetic field.
In section IV, we derive exact expressions of spin Hall conductivity 
for the generic systems and show the relation between SHC and
the spin-orbit field.
We provide a summary of our work in section V.

\sn{Generic spin-orbit coupled two-dimensional fermionic systems}
The Hamiltonian of a single fermion of mass $m$ and charge $q$ 
in a two-dimensional fermionic system with a generic SOI \cite{generic-H}
in III-V heterojunctions 
is given by
\bearr \label{hamil-1}
H & = & \dfrac{\hbar^2k^2}{2m} 
+ \dfrac{i\alpha_l}{2\hbar^l}\lf(p_-^l\sigma_+ - p_+^l\sigma_-\rt), 
\eearr
where $l=1,2,3 $ and $ \alpha_l $ is the spin-orbit coupling constant
whose dimension varies with $l$.
$ p_{\pm} = p_x \pm i p_y $ and 
$ \sigma_{\pm} = \sigma_x \pm i\sigma_y $ are the complex representation of
the momentum operators and Pauli spin matrices, resp.
When $l=1$, the Hamiltonian represents two dimensional electron gas with 
the $k$-linear Rashba \cite{rashba,rashba1} or Dresselhaus spin-orbit interaction 
\cite{dresselhaus-55}.
The spin-orbit interaction corresponding to $l=2$ arises when an in-plane
magnetic field is applied to the 2D heavy hole gas \cite{generic-H,bula,book} 
formed at the GaAs heterojunctions.
In this case the spin-orbit coupling constant varies linearly with the applied
magnetic field i.e. $ \alpha_2 \propto B $.
Therefore, $k$-quadratic spin-orbit interaction is invariant under
the time-reversal operation. 
The $k$-quadratic term is dominating in the high symmetry growth directions [001] and [111] 
of the heavy holes in GaAs heterojunctions.
For $l=3$, the Rashba spin-orbit interaction is cubic in momentum 
\cite{winkler,winkler1,chesi}. The $k$-cubic Rashba spin-orbit
interaction is present in 2D heavy hole gas as well as on the surface of
SrTiO${}_3$.
Typical values of the spin-orbit coupling constant $\alpha_l$ \cite{book} are
$\alpha_1 \simeq 10^{-11}-10^{-13}$ eV-m, $\alpha_2 \simeq 10^{-20} $ eV-m${}^2$ for 
$B \sim 1 $ T and width of the quantum well $W = 2 \times 10^{-5} $ m, 
and $ \alpha_3 = 10^{-27}-10^{-28} $ eV-m${}^3$.
We note in passing that Dirac materials (e.g. graphene and graphene-like materials, 
topological insulators etc.) have a different form of spin-orbit interaction, 
which has not been taken up in this investigation.

The eigenvalues are 
$ \veps_\lambda = \hbar^2k^2/(2m) + \lambda \alpha_l k^l $ 
and the corresponding normalized eigenvectors are
\be
|\chi^{\lambda}\ra = \cfrac{e^{i {\bf k} \cdot {\bf r}}}{\sqrt 2} 
\bpm 1 \\ -\lambda ie^{i l \theta} \epm,
\ee
where $ \lambda = \pm $ denote two spin-split energy branches and
$ \theta = \tan^{-1}(k_y/k_x)$.
For the given Fermi energy $\veps_F$,
the above two energy branches give rise to two different Fermi
wave vectors $k_F^\pm$ fulfilling
\be
\label{fermi-en}
\veps_F = \cfrac{(\hbar k_F^\pm)^2}{2m} \pm \alpha_l (k_F^\pm)^l \ee
with $k_F^+<k_F^-$  for positive $\alpha_l$. Eq.(\ref{fermi-en}) actually
contains two equations; subtracting one from the other gives
\be
\label{en-diff}
\cfrac{\hbar^2}{2m} \lf( (k_F^+)^2 - (k_F^-)^2 \rt)
+ \alpha_l \lf( (k_F^+)^l + (k_F^+)^l \rt) = 0 . \ee
The total carrier density $n_F$ is given by
\be
\label{hole-density} 
n_F = \cfrac{1}{4\pi}\lf[(k_F^+)^2 + (k_F^-)^2\rt].
\ee
One can easily get $ k_F^\pm$ for different values of $l$ by
solving \cite{loss-05} the above 
two equations. These are given by
\begin{align*}
k_{F}^{\pm} &=  \sqrt{2\pi n_F - q_{1}^2} \mp q_1 & l &= 1 \\
&= \sqrt{2\pi n_F\lf (1 \mp 2 q_2 \rt)} & l &= 2 \\
&= \sqrt{3\pi n_F - \frac{L_F}{8 q_3^2} }  \mp \frac{L_F}{4q_3} & l &= 3 ,
\end{align*}
where 
$ L_F = \lf[1 - \sqrt{1-  16 \pi n_F q_3^2}\rt] $ with 
$ q_l = m \alpha_l/\hbar^2 $. Note that the dimension of $ q_l $
is $ L^{l-2} $.

\sn{Lorentz-like Spin-Orbit Force and spin-orbit field}

In this section we derive 
the spin-orbit force from the generic spin-orbit interaction Hamiltonian.
Using the Heisenberg equation of motion \cite{eugen}, $ i \hbar \dot A = [A,H] $,
we derive, by turn,
\be \label{vel1}
\dot {\nvec r} = \dfrac{\nvec p}{m} + \dfrac{l \alpha_l}{2\hbar^l}
\Big[p_-^{l-1}\sigma_+ (i \hvec x+\hvec y)
- p_+^{l-1}\sigma_- (i \hvec x-\hvec y) \Big]
\ee
and 
\begin{align} \label{acc}
\ddot {\nvec{r}} & = 
\dfrac {2 l \alpha_l^2}{\hbar ^{2l+1} } 
p^{2(l-1)}  (\nvec{p} \btimes \hvec{z})  \sig_z.
\end{align}
We wish to express Eq.(\ref{acc}) in a form similar to the Lorentz force
[$ {\bf F}_{\rm L} = q (\dot{\nvec{r}} \btimes {\bf B})$] acting on a charge
particle $q$ by the external magnetic field ${\bf B}$. 

To do so, we need to know the relation between 
momentum and velocity for the generic spin-orbit interaction Hamiltonian.
One can easily verify that 
\begin{align}
\dfrac {\nvec p \btimes \hvec z}{m} & = \dot{\nvec{r}} \btimes \hvec{z} 
- \dfrac{l \alpha _l}{\hbar^l} p^{l-1} \nn \\
\label{vel} & \times 
\lf\{ \nvec \sig \cos[(l-1)\theta] + (\nvec \sig \btimes \hvec z)
\sin[(l-1) \theta] \rt\}.
\end{align}

Putting Eq.(\ref{vel}) into Eq. (\ref{acc}) and rewriting it in the form of
the Lorentz force $ {\bf F}_{\rm L}$, we have
\begin{align} \label{sof-g}
{\bf F}_{\rm so} & = m \ddot {\nvec{r}} =
e \Big[ \dot{\nvec{r}} \btimes  
\dfrac {2 l m^2 \alpha_l^2}{e \hbar ^{2l+1} }  p^{2(l-1)}\sig_z \hvec z \Big] 
- \dfrac{i l m \alpha_l}{\hbar^l} p^{l-1} \nonumber \\
& \times  
\Big[ \nvec \sig \cos[(l-1)\theta] 
+ (\nvec \sig \btimes \hvec z)  \sin[(l-1) \theta] \Big] 
\btimes \hvec z .
\end{align}
This is one of the main results. 
The first term on the right hand side of the above equation
is exactly similar in form to
the Lorentz force ${\bf F}_{\rm L}$ and one can immediately identify that the
spin-orbit interaction dependent {\it pseudo magnetic field }  operator is 
\be
{\bf B}_{\rm so}^{(l)} = 
\dfrac {2 l m^2 \alpha_l^2}{e \hbar ^{2l+1} }  p^{2(l-1)} \hvec z 
\otimes \sigma_z .
\ee
Note that the pseudo magnetic field depends on the
two-dimensional momentum operator ($p$) as well as on the 
Pauli spin operator $\sigma_z$. 
It is perpendicular to the plane of the system. The Pauli spin matrix
$\sigma_z $ appearing in ${\bf B}_{\rm so}^{(l)} $ will act on 
spinor of the injected charge carriers.
The physical significance of the last term
of Eq. (\ref{sof-g}) remains unclear.
However, the last term does not contribute to the calculation of
the average force acting on the charge carriers.

Now we would like to calculate magnitude of the spin-orbit field which
can be obtained from the above equation.
The spin-orbit field produced by a single fermion with the wave vector $k$ is
then
\be \label{sofield}
 B_{\rm so}^{(l)}(k) =
\dfrac {2 l m^2 \alpha_l^2}{e \hbar ^{3} }  k^{2(l-1)}.
\ee

The spin-orbit field does not depend on the carrier density for
the case $l=1 $. However, it depends on the 
charge carrier density for the cases $l=2 $ and $l=3 $.
For $k$-linear spin-orbit interaction, the spin-orbit field produces by
each electron is  
$ B_{\rm so}^{(1)} (k) = 2 m^2 \alpha_1^2/(e\hbar^3) \sim 10^{-4} $ T
for $ \alpha_1 = 10^{-12} $ eV-m and $ m = 0.04 m_e $. 
Similarly, the spin-orbit field produced by the hole at the
Fermi surface are
$ B_{\rm so}^{(2)} (k) \sim 4\pi n_F m^2 \alpha_1^2/(e\hbar^3)
\sim  10^{-3} $ T for $ \alpha_2 = 10^{-19} $ eV-m${}^2$ and
$ B_{\rm so}^{(3)} (k) \sim 2 m^2 \alpha_3^2 k_F^4/(e\hbar^3)   
\sim 10^{-2} $ T for $\alpha_3 = 10^{-27} $ eV-m${}^3$. 
For the estimate of $\alpha_2 $ and $\alpha_3 $, we have used
$ m = 0.4 m_e $ and $ n_F = 10^{14}$ m${}^{-2}$. 
It shows that the spin-orbit field produced by a single fermion at
the Fermi surface is quite strong.

In the spin Hall effect experiment,  
suppose we inject unpolarized (equal number of spin-up and
spin-down charge carriers) charge carriers along the $x$ axis. 
The spin-orbit force will act on the
injected unpolarized charge carriers due to the spin-orbit field 
produced in the $z$ direction by all the charge carriers.
The spin-up and spin-down electrons feel spin-orbit force 
along $\mp \hat y$, respectively,
resulting in a spin separation across $y$ direction.
Thus this spin-orbit field certainly explains SHE.

\sn {Relation between spin Hall conductivity and spin-orbit field}
In this section we shall derive the spin Hall conductivity 
in 2D fermionic systems with generic spin-orbit interaction.
Note that the generic SOI (last term of Eq. (\ref{hamil-1})) can 
be rewritten in the form of Zeeman interaction as 
$ H_R = (gq)/(2m){\bf B}_{\rm Z}({\bf k}) \cdot {\bf S} $,  
where $ {\bf S} = J \hbar {\bs \sigma} $ is the total
angular momentum operator of the carriers and the wave vector 
dependent pseudo Zeeman field is given by
\be
{\bf B}_{\rm Z}({\bf k}) =  \frac{2m \alpha_l}{ J g q \hbar} k^l
\Big[ \sin(l\theta) \hat x - \cos(l\theta) \hat y\Big] 
\ee 
This field is responsible for the spin-splitting even in the absence of
external magnetic fields.
For electrons in n-type heterojunction, $ J =1/2$ and
for heavy holes in p-type heterojunctions, $J =3/2$.

In presence of a weak electric field ${\bf E} = E \hat x $,
the equation of motion of a charge carrier 
is given by
\be \label{eqmE}
\cfrac{d {\bf p}}{dt} = q {\bf E} 
- \cfrac{{\bf p} - {\bf p}_0}{\tau},
\ee
where $ 1/\tau$ is the impurity scattering rate, 
$ {\bf p} = \hbar {\bf k} $ is the momentum  at time $t$ and 
$ {\bf p}_0 = \hbar {\bf k}_0$ is the initial momentum in 
absence of the external electric field.
For convenience, we first assume an ac 
electric field and in the end we will take the dc limit. Assume a 
weak ac electric field  $ {\bf E} = E_x e^{i \omega t} \hat x$.
The solution of Eq. (\ref{eqmE}) is given by
\be \label{wavevector}
k_x(t) = k_{0x} + \frac{(qE_x/\hbar)e^{i \omega t}}{i \omega + 1/\tau},
\h{1cm} k_y(t) = k_{0y}. \ee

The Heisenberg equation of motion of the spin vector ${\bf S} $ is given by
\be
\frac{\mrm d {\bf S}}{\mrm dt}
= \frac{g q J {\bf B}_{\rm Z}}{m} \times {\bf S},
\ee
which, on simplification, yields
\be \label{h-eom-2}
\bpm \dot S_x \\ \dot S_y \\ \dot S_z \epm
= \frac{2 \alpha_l k^l}{\hbar}
\bpm 0 & 0 & -\cos {(l \theta)} \\ 0 & 0 & -\sin {(l \theta)} \\ 
\cos {(l \theta)} & \sin {(l \theta)} & 0 \epm
\bpm S_x \\ S_y \\ S_z \epm ,
\ee
where $ \dot S_j $ denotes time derivative of $S_j$ with 
$ j=x,y,z$.
%\be \label{h-eom-2}
%\cfrac{\mrm d}{\mrm dt} \bpm S_x \\ S_y \\ S_z \epm
%= \frac{2 \alpha_l k^l}{\hbar}
%\bpm 0 & 0 & -\cos {(l \theta)} \\ 0 & 0 & -\sin {(l \theta)} \\ 
%\cos {(l \theta)} & \sin {(l \theta)} & 0 \epm
%\bpm S_x \\ S_y \\ S_z \epm .
%\ee

The pseudo Zeeman field felt by the charge carrier will become
time-dependent due to the ac electric field. The charge carrier's spin
will precess around the equilibrium orientation periodically with time.
In the linear response regime, the dynamic precession of the spin
of a particle around the equilibrium orientation can be expressed
as 
\be \label{fluc}
S_j^{\lambda}({\bf k},t) 
= S_j^{(0), \lambda}({\bf k}_0)
+ \Omega_j^{\lambda} e^{i \omega t},
\ee
where $ S_j^{(0), \lambda}({\bf k}_0)
=J \hbar \la {\bf k}_0 \lambda | \sigma_j |
{\bf k}_0 \lambda \ra $ is the expectation value of the spin operator 
which is initially in the spinor eigenstate
$ | {\bf k}_0, \lambda \ra $ before the external electric field is applied.
Also, $\Omega_j^{\lambda} $ is the amplitude of the deviation of
the spin from the equilibrium state under the action of
the electric field.
Within the linear approximation, simplification of the Heisenberg 
equations of motion yields
\be
\Omega_z^{\lambda} = - \cfrac{2 \lambda l \alpha_l J \hbar k_0^{l-2}k_{0y} }
{\hbar^2 \omega^2 - 4 \alpha_l^2 k_0^{2l} }
\h{4 pt}\cfrac{i \omega q E_x}{i \omega + 1/\tau}\h{6 pt} ,
\ee
\be
\Omega_x^{\lambda}
= \cfrac{4 \lambda l \alpha_l^2 J k_0^{2l-2} k_{0y}\cos{(l\theta_0)} }
{\hbar^2 \omega^2 - 4 \alpha_l^2 k_0^{2l}  }
\h{4 pt}\cfrac{q E_x}{i \omega + 1/\tau}
\ee
and
\be
\Omega_y^{\lambda}
= \cfrac{4 \lambda l \alpha_l^2 J k_0^{2l-2} k_{0y}\sin{(l\theta_0)} }
{\hbar^2 \omega^2 - 4 \alpha_l^2 k_0^{2l}  }
\h{4 pt}\cfrac{q E_x}{i \omega + 1/\tau} \h{4pt}.
\ee
Note that the out-of-plane spin fluctuation ($\Omega_z^{\lambda} $) arises due to
the applied in-plane electric field, which is linear in $\alpha_l$.
On the other hand, the in-plane spin fluctuations are quadratic in $\alpha_l$.

The conventional spin current operator associated with $z$-polarized spin 
moving in the $y$-direction is given by 
$ \hat J_y^{z}= J \hbar (v_y \sigma_z + \sigma_z v_y)/2  $, where 
the $y$-component of the velocity operator $v_y $ is obtained 
from Eq. (\ref{vel1}). After simplification, it becomes
$ \hat J_y^{z}= J \hbar (\hbar k_y/m) \sigma_z = (\hbar k_y/m)S_z $.
Note that the spin current operator does not depend on $\alpha$ explicitly. 
The spin current density at zero temperature is
$ J_y^z = \sum_{\lambda} \int \frac{\mrm d^2k}{(2\pi)^2}
S_z^{\lambda} ({\bf k}) (\hbar k_{y}/m) 
= \sigma_{\rm sH}^{(l)} (\omega,\tau) E_x $, 
where the spin Hall conductivity is given by
\be
\sigma_{\rm sH}^{(l)} (\omega,\tau) =
\sum_{\lambda} \int \frac{\mrm d^2k}{(2\pi)^2} 
\frac{2 \lambda l \alpha_l q J \hbar^2 k^{l-2} k_{y}^2}
{m\lf (4 \alpha^2 k^{2l} - \hbar^2 \omega^2 \rt)} 
\frac{i \omega}{i \omega + 1/\tau}.
\ee
In presence of impurity, $1/\tau \neq 0 $ and hence in the
dc limit $\omega =0$, one can see that the spin Hall conductivity
$\sigma_{\rm sH} $ vanishes exactly. It shows that even a small
amount of impurity can destroy the spin Hall conductivity completely
\cite{disorder,disorder1}.

We consider an infinite system without any impurity ($1/\tau=0$)
and then taking dc limit ($\omega \rightarrow 0$), we get . 
\be \label{shc0-1}
\sigma_{\rm sH}^{(l)}
= \frac{l q J \hbar^2}{8 \pi m \alpha_{l}}
\int_{k_F^-}^{k_F^+} \frac{\mrm dk}{k^{l-1}}.
\ee
%Here we shall present exact results of the spin Hall conductivity 
%of different systems with no impurities i.e. $ 1/\tau = 0 $. 
%For $l=1$ ($q=-e, J = 1/2$), one can easily get
%$ \sigma_{\rm sH}^{(1)} = e/(8 \pi) $. 
%This result is exactly same as obtained in Ref. \cite{niu-04-2}.
For $l=1$ ($q=-e, J = 1/2$), we simply reproduce
the result $ \sigma_{\rm sH}^{(1)} = e/(8 \pi) $, exactly
same as obtained by Niu \textit{et al}\cite{niu-04-2}.
For $l=2$ ($q=e, J=3/2$), we have 
\bearr
\sigma_{\rm sH}^{(2)} & = & \frac{3 e \hbar^2}{8 \pi m \alpha_2} 
\ln{\Big(\frac{k_F^+}{k_F^-}\Big)} \nn \\
& = &
-\frac{3e}{4\pi} \lf(1 + \frac{4}{3} q_2^2 
+ \frac{16}{5} q_2^4 + \cdots \rt) \nn
\eearr
where $ q_2 = m \alpha_2/\hbar^2 $.
It increases logarithmically with $ k_F^+/k_F^-$.
For $l=3$ ($q=e, J=3/2$), we have 
\bearr 
\sigma_{\rm sH}^{(3)}  
& = & \frac{9e \hbar^2}{16 \pi m \alpha_3}
\Big(\frac{k_F^+ - k_F^-}{k_F^+ k_F^-}\Big) 
\nn \\
& = & -\frac{9e}{8 \pi} 
\lf(1 +  8 \pi n_F q_3^2 
+96 \pi^2 n_F^2 q_3^4  + \cdots \rt) \nn. 
\eearr
Here again the zeroth order term matches with the value
reported by Loss \textit{et al}\cite{loss-05}.
The above two series expansions are valid since 
$ q_2 \ll 1 $ and $ q_3^2 n_F \ll 1 $ for the typical parameters
in various systems.

Using Eq. (\ref{sofield}), Eq. (\ref{shc0-1}) can be re-expressed 
%in terms of the $B_{\rm so}^{(l)} $ 
%Equation (\ref{shco-1}) can be re-written 
as
\be
\sigma_{\rm sH}^{(l)}
= \frac{q}{8\pi} J \sqrt{\frac{l^3}{\pi}} 
\int_{k_F^-}^{k_F^+} \sqrt{ \frac{\phi_0 }{ B_{\rm so}^{(l)}(k) }} \mrm dk,
\ee
where $ \phi_0 = h/e $ is the unit of magnetic flux quanta.
Two important conclusions can be drawn from the above equation:
i) it is directly related to the flux quanta $\phi_0$ although no real magnetic field
is applied, and ii) it varies with inverse square root of the 
spin-orbit field $B_{\rm so}^l(k) $, similar to the charge Hall conductivity
\be
\sigma_H = \frac{e n_F}{B} = \frac{e^2}{h} n_F\frac{\phi_0}{B},
\ee
which is inversely proportional to the external magnetic field $B$.

%It is remarkable that not only the results obtained by
%this semi-classical approach are consistent with the zeroth order
%terms obtained by rigorous quantum approaches (for the cases
%$l=1$ and $l=3$), but
%the expressions obtained are also exact and complete
%with all the higher order terms, while the rigorous approaches
%are unable to give higher order terms. This may be considered
%as a success in view of simplicity, consistency and exactness.
%Also, we obtain completely new result for the case $l=2$.
%
%The SHC for $l=1$ is universal in a sense that it is independent
%of $n_F$ as well as on the spin-orbit coupling constant. On the other hand,
%the SHC for $l=2$ depends on SOC constant only and for $l=3$ 
%it depends on the combination of $n_F$ and $\alpha_3$. 
%Therefore, the SHC in general is not universal since it depends on the system
%parameters for $l=2 $ and $l=3 $ cases.

\sn{Summary}

In this work, we have derived the correct Lorentz-like spin-orbit
force with the associated spin-orbit field of 
two-dimensional fermionic systems
with generic spin-orbit interaction. The spin-orbit field is
directed normal to the plane and its magnitude depends on the
density as well as spin-orbit coupling constant. 
We have presented a closed-form expression of the spin Hall conductivity 
in a generic spin-orbit coupled system.
Furthermore, we have found that the spin Hall conductivity is 
inversely proportional to the square root of the spin-orbit field. 
Moreover, it is directly related
to the flux quanta although no real magnetic field is applied.

%We found that although spin Hall conductivity for 2DEG is universal but
%it depends on the system parameters
%such as density and spin-orbit coupling constant for heavy hole systems. 
%One can conclude that
%spin Hall conductivity is not an universal quantity.


\begin{thebibliography}{55}

\bibitem {ashcroft}
N. W. Ashcroft, and N. D. Mermin,
Solid State Physics (Harcourt, Orlando, 1976)
  
\bibitem {zhang-03}
S. Murakami, N. Nagaosa, and S. C. Zhang,
Science {\bf 301}, 1348 (2003).
  
\bibitem {zhang-04-1}
S. Murakami, N. Nagaosa, and S. C. Zhang,
Phys. Rev. B {\bf 69}, 235206 (2004).
  
\bibitem {niu-04-2}
J. Sinova, D. Culcer, Q. Niu, N. A. Sinitsyn,
T. Jungwirth, and A. H. MacDonald,
Phys. Rev. Lett. {\bf 92}, 126603 (2004).
  
\bibitem {loss-04}
J. Schliemann and D. Loss,
Phys. Rev. B {\bf 69}, 165315 (2004).

\bibitem {sinova-04}
N. A. Sinitsyn, E. M. Hankiewicz, W. Teizer, and J. Sinova,
Phys. Rev. B {\bf 70}, 081312(R) (2004).

\bibitem{perel-71-1}
M. I. Dyakonov and V. I. Perel,
Sov. Phys. JETP Lett. {\bf 13}, 467 (1971).

\bibitem{perel-71-2}
M. I. Dyakonov and V.I. Perel, 
Phys. Lett. A {\bf 35}, 459 (1971).


\bibitem{hirsch-99}
J. E. Hirsch,
Phys. Rev. Lett. {\bf 83}, 1834 (1999).

\bibitem{ahe}
P. Noziére and C. Lewiner, 
J. Phys. {\bf 34}, 901 (1973).

\bibitem{ahe1}
N. A. Sinitsyn,
J. Phys.: Condens. Matter 20, 023201 (2008).


\bibitem{kato-04}
Y. Kato, R. C. Myers, A. C. Gossard, and D. D. Awschalom, 
Science {\bf 306}, 1910 (2004).

\bibitem{sinova-05}
J. Wunderlich, B. Kaestner, J. Sinova, and T. Jungwirth,
Phys. Rev. Lett. {\bf 94}, 047204 (2005).

\bibitem{murakami-04}
S. Murakami,
Phys. Rev. B, {\bf 69}, 241202(R) (2004).

%\bibitem{murakami-04}
%\Name{Murakami S.}
%\Review{Phys. Rev. B} {69} {2004} {241202(R)}

\bibitem {berenvig}
B. A. Bernevig, J. P. Hu, E. Mukamel, and S. C. Zhang,
Phys. Rev. B, {\bf 70}, 113301 (2004).

%\bibitem {berenvig}
%\Name{Bernevig B. A., Hu J. P,, Mukamel E., \and Zhang S. C.,}
%\Review{Phys. Rev. B} {70} {2004} {113301}

\bibitem{shen-04} %Berry phase and SHE
S. Q. Shen,
Phys. Rev. B, {\bf 70}, 081311(R) (2004).

%\bibitem{shen-04} %Berry phase and SHE
%\Name {Shen S. Q.}
%\REVIEW {Phys. Rev. B} {70} {2004} {081311(R)}

\bibitem{culcer}
D. Culcer, J. Sinova, N. A. Sinitsyn, T. Jungwirth,
A. H. MacDonald, and Q. Niu,
Phys. Rev. Lett., {\bf 93} 046602 (2004).

%\bibitem{culcer}
%\Name{Culcer D., Sinova J., Sinitsyn N. A., Jungwirth T.,
%MacDonald A. H., \and Niu Q.}
%\Review{Phys. Rev. Lett.} {93} {2004} {046602}

\bibitem{loss-05}
J. Schliemann and D. Loss,
Phys. Rev. B {\bf 71}, 085308 (2005).


\bibitem{huang-06}
L. Hu and Z. Huang,
Physics Letters A {\bf 352}, 250 (2006).

\bibitem{ijmpb}
J. Schliemann,
Int. J. Mod. Phys. B, {\bf 20}, 1015 (2006).

%\bibitem{ijmpb}
%\Name{ Schliemann J.}
%\Review{Int. J. Mod. Phys. B}{20} {2006} {1015}

\bibitem{eugen}
E. M. Chudnovsky, 
Phys. Rev. Lett. {\bf 99}, 206601 (2007).

\bibitem{slf}
I. V. Tokatly,
Phys. Rev. Lett. {\bf 101}, 106601 (2008).

\bibitem{njp}
T. Fujita, M. B. A. Jalil, and S. G. Tan,
New J. Phys. {\bf12}, 013016 (2010).

\bibitem{berry}
S.-Q Shen,
Phys. Rev. B {\bf 70}, 081311(R) (2004).


\bibitem{sof1}
J. Li, L. Hu, and S. Q. Shen,
Phys. Rev. B {\bf 71}, 241305(R) (2005).

\bibitem{sof2}
B. K. Nikolic, L. P. Zarbo, and S. Welack,
Phys. Rev. B. {\bf 72}, 075335 (2005).

\bibitem{generic-H}
S. Chesi and G. F. Giuliani,
Phys. Rev. B {\bf 75}, 155305 (2007).


\bibitem {rashba}
E. I. Rashba,
Fiz. Tverd. Tela (Leningrad) {\bf 2}, 1224 (1960)
[Sov. Phys. Solid State {\bf 2}, 1109 (1960)]

\bibitem {rashba1}
Y. A. Bychkov and E. I. Rashba,
J. Phys. C {\bf 17}, 6039 (1984).

\bibitem {dresselhaus-55}
G. Dresselhaus,
Phys. Rev. {\bf 100}, 580 (1955).

\bibitem{bula}
D. V. Bulaev and D. Loss, 
Phys. Rev. Lett. {\bf 98}, 097202 (2007).

\bibitem {book}
W. Winkler,
Spin-Orbit Coupling Effects in Two-Dimensional Electron and 
Hole Systems (Springer Verlag, 2003)

%\bibitem{book}
%\Name{Winkler W.}
%\Review{Spin-Orbit Coupling Effects in Two-Dimensional Electron and 
%Hole Systems (Springer Verlag-2003)}

\bibitem{winkler}
R. Winkler,
Phys. Rev. B {\bf 62}, 4245 (2000).

\bibitem{winkler1}
R. Winkler, H. Noh, E. Tutuc, and M. Shayegan,
Phys. Rev. B {\bf 65}, 155303 (2002).

\bibitem{chesi}
S. Chesi, G. F. Giuliani, L. P. Rokhinson, L. N. Pfeiffer,
and K. W. West, 
Phys. Rev. Lett. {\bf 106}, 236601 (2011).


\bibitem{disorder}
S. I. Erlingsson, J. Schliemann, and D. Loss,
Phys. Rev. B {\bf 71}, 035319 (2005).

\bibitem{disorder1}
Ol’ga V. Dimitrova,
Phys. Rev. B {\bf 71}, 245327 (2005).





\end{thebibliography}
\end{document}